\documentclass{elsart4}

 \usepackage{graphicx}

\usepackage{amssymb}

\begin{document}

\begin{frontmatter}

\title{A critical discussion of calculated modulated structures, Fermi surface
  nesting and phonon softening in magnetic shape memory alloys
  Ni$_2$Mn(Ga, Ge, Al) and Co$_2$Mn(Ga, Ge)}

\author{A. T. Zayak\corauthref{cor1}}
\ead{alexei@thp.uni-duisburg.de}
\corauth[cor1]{Tel: +49-203-3761606 ; fax:  +49-203-3763665 }
\author{P. Entel}
\address{Institute of Physics, University of Duisburg-Essen,
             Duisburg Campus, 47048 Germany}

%%\title{}


%\author{}
%\address{}

\begin{abstract}
A series of first principles calculations have been carried out in order to
discuss electronic structure, phonon dynamics, structural
instabilities and the nature of martensitic transformations of the Heusler 
alloys Ni$_2$Mn(Ga, Ge, Al) and Co$_2$Mn(Ga, Ge). The calculations show that
besides electronic pecularities like Fermi--surface nesting,
hybridizing optical and acoustic phonon modes are important for the
stabilization of the modulated martensitic structures.
\end{abstract}

%%%%%%%%%use  the \KEY command at the begin of keyword text%%%%%%%%%
\begin{keyword}
\PACS %code\sep code
      61.66.Dk \sep 63.20.Dj \sep 64.70.Rh
\KEY  Heusler alloys\sep Phonon instabilities \sep Incommensurate phase
\end{keyword}
\end{frontmatter}

\section{Introduction}\label{Int}

%%%main text


In this paper, we discuss 
microscopic details of the structural instabilities observed in 
ferromagnetic Heusler alloys like the magnetic shape-memory (MSM) system
Ni-Mn-Ga \cite{Ullakko-96}.
These alloys are of technological importance being used for
mechanical applications based on specific elastic properties of the
martensites. Due to ferromagnetic order magnetic fields 
allow to control the crystal structure. Ni$_2$MnGa has now become a reference 
system for all 
investigations related to the MSM technology. It has been shown
that in moderate fields of the order of 1 T, the structural
deformations in Ni$_2$MnGa can reach $\sim$10\% \cite{Sozinov}. This makes the
MSM technology important for new kinds of micro-mechanical
sensors and actuators \cite{Wei-00}.
Serious attemps are undertaken to search for new systems which would show
even better MSM performance \cite{Takeuchi}.

\begin{figure} % Fig. 1
  \begin{center}
  \includegraphics*[angle=0,width=8cm]{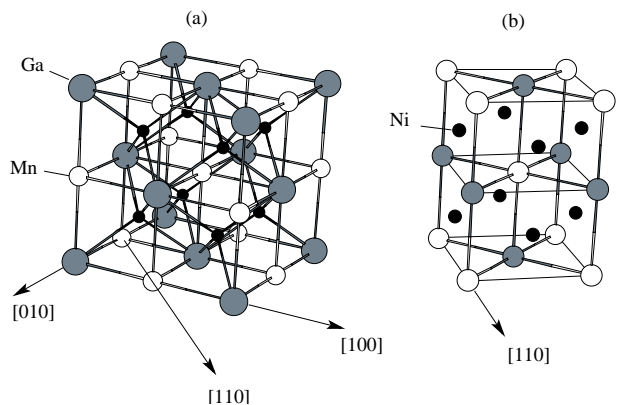}
  \end{center}
  \caption{(a) The cubic L2$_1$ structure of the Heusler alloy 
  Ni$_2$MnGa. The tetrahedrally coordinated structure of Ga around Ni
  is clearly visible. Mn also forms a
  tetrahedral coordination, but in case of Ga the $p$-electrons are able to
  couple covalently to the Ni atoms. (b) Reduced tetragonal structure used
  in the calculations. }
  \label{struc}
\end{figure}

%Thus, about 6\% strain was obtained in Ni-Mn-Ga
%for the 10M martensitic structure ($a = b \ne c$,\ \ $c/a = 0.94$) with
%superposed five-layered modulation. This modulation has a shape of a static
%wave along the [110] direction with a
%polarization  vector along [1\=10], i.e., the shuffling of the atoms is
%in-plane  \cite{Martynov-92,Zayak-03}. The value of 6\% has 
%been predicted as maximum value for the 10M structure. Recently, new
%achievements have been reported where 10\% magnetic-field-induced strains were
%obtained in Ni-Mn-Ga \cite{Sozinov}. In this case the martensite has the 
%14M structure with orthorhombic symmetry and superposed
%7-layer modulation \cite{Pons-Stack}.

%\begin{figure} % Fig. 1
%  \begin{center}
%  \includegraphics*[angle=0,width=6.5cm]{Ni2MnGa_ohne.eps}
%  \end{center}
%  \caption{Calculated phonon dispersions for the cubic L2$_1$ structure of
%  Ni$_2$MnGa.}
%  \label{nimnga}
%\end{figure}

The MSM effect is a completely material dependent property related to
the different kinds of martensite formed at corresponding working
conditions. 
A complete microscopic explanation of the martensitic modulated structures is
not available which, however, is the key to understand the different properties
of apparently similar Heusler alloys like Ni$_2$MnGa and Co$_2$MnGa. 
So far, one has tentatively related the modulated shuffling of the
atoms and the phonon softening to the specific shape of the
Fermi surface of Ni$_2$MnGa \cite{Harmon,Claudia}. In this way the Kohn
anomaly driven by the Fermi--surface  
nesting can lead to the structural instabilities of the
crystal. Unfortunately this nesting picture does not give a comprehensive
understanding of the micro-mechanics involved in the shuffling of the
atoms on the atomic scale.
It also does not allow a complete characterization of the expected
strong coupling of the electrons and phonons. Finally atomic disorder,
important for the stabilization of the modulated structures, might
diminish the impact of nesting features. 

With the aim to connect the electron and phonon properties of the Heusler
alloys we have carried out a series of \textit{ab-initio}
investigations. Phonon dispersions of 
several Heusler alloys have been calculated in order to discuss differences
between stable and unstable structures.

\begin{figure} % Fig. 4
  \begin{center}
  \includegraphics*[angle=0,width=6.2cm]{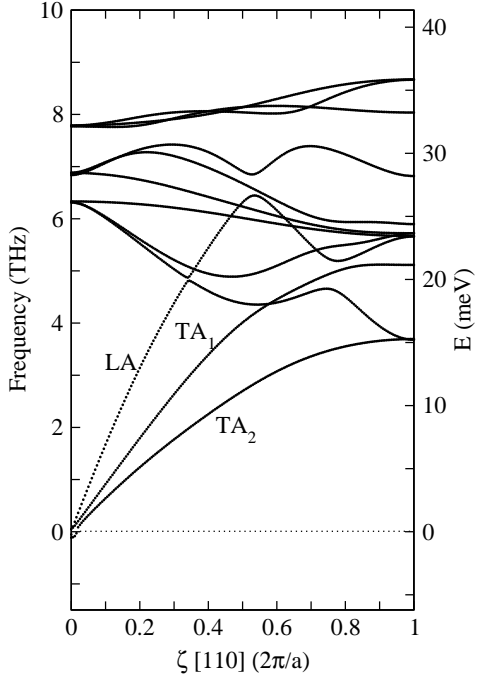}
  \end{center}
  \caption{Calculated phonon dispersions for the cubic L2$_1$ structure of
  Co$_2$MnGe.}
  \label{comnge}
\end{figure}

%%%%%%%%%%%%%%%%%%%%%%%%%%%%%%%%%%%%%%%%%%%%%%%%%%%%%%%%%%%%
\section{Computational details}
%%%%%%%%%%%%%%%%%%%%%%%%%%%%%%%%%%%%%%%%%%%%%%%%%%%%%%%%%%%%

In order to calculate the phonon dispersions we have used the direct
force--constant method \cite{Parlinski,Parlinski_M}.
This method uses the forces
calculated via the Hellmann-Feynman theorem in the total energy
calculations for a supercell with periodic boundary
conditions, which define to the corresponding dynamical matrix.
The method works within the
harmonic approximation and the accuracy of it
strongly depends on the size of the considered supercell \cite{Parlinski}.

\begin{figure} % Fig. 3
  \begin{center}
  \includegraphics*[angle=0,width=6.0cm]{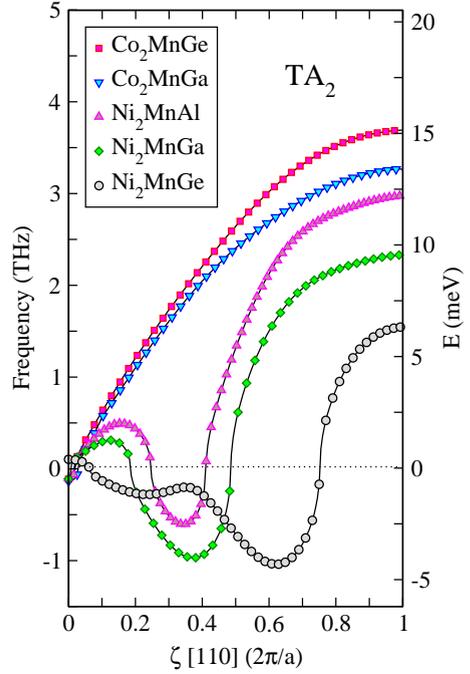}
  \end{center}
  \caption{Dispersion of the transverse acoustic phonon modes calculated for
  different alloys: Ni$_2$MnGa, Ni$_2$MnGe, Ni$_2$MnAl, Co$_2$MnGa and
  Co$_2$MnGe. Positions of the softening for Ni$_2$MnGa and Ni$_2$MnAl
  coinside which we relate to the same valence electron concentration of these
  alloys ($e/a = 7.5$). Increasing of $e/a$ shifts the softening to larger
  $\zeta$, while decreasing $e/a$ shifts the softening towards the center
  of the Brillouin zone where Co$_2$MnGa and Co$_2$MnGe can be considered as
  limiting cases.}
  \label{modes}
\end{figure}

The Vienna {\it Ab-initio} Simulation Package (VASP)
\cite{Kresse-96} has been used to perform the electronic
structure calculations. The projector-augmented wave formalism (PAW),
implemented in this package \cite{PAW}, leads to very accurate
results compared to all-electron methods. The electronic exchange and 
correlation are treated by using the
generalized gradient approximation.

All alloys considered in this work exist in the cubic L2$_\mathrm{1}$
structure shown in Fig. \ref{struc}(a)  (in case of Ni$_2$MnGe the
L2$_\mathrm{1}$ structure has not been reported so far). 
All directions are refered to the cubic
structure. We take five tetragonal cells shown in Fig. \ref{struc}(b) and merge
them together along the [110] direction giving a
$1 \times 5 \times 1$ orthorhombic supercell. The five
tetragonal unit cells allow for five \textit{k}-points
in the Brillouin zone along the [110] directions for which the calculated
phonon frequencies will be exact \cite{Parlinski}.

\begin{figure} % Fig. 4
  \begin{center}
  \includegraphics*[angle=0,width=5.8cm]{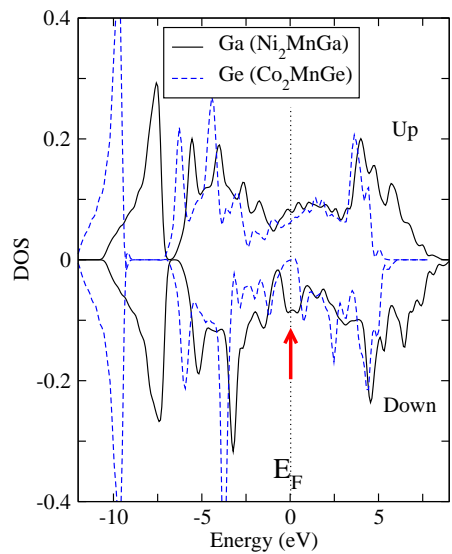}
  \end{center}
  \caption{Site projected electronic density of states of Ni$_2$MnGa
  and Co$_2$MnGe showing the difference in the spin-down electronic states
  close to the Fermi level of stable and unstable
  Heusler alloys. In Ni$_2$MnX alloys (X = Ga, Ge, Al, In) the peak
  marked by the
  arrow arises from $p$-states of the X atom, whereby the double peak
  structure is hybridization splitting. The X-projected
  $p\downarrow$-orbitals 
  are coupled to the Ni-projected $d\downarrow$ states. We tentatively argue
  that this hybridization will be influence by the electron-phonon interaction.}
  \label{dos}
\end{figure}

\section{Discussion}

The calculated phonon dispersions for cubic Co$_2$MnGe 
are shown in Fig. \ref{comnge}. Further dispersions not shown
here have been obtained for Ni$_2$MnGa, 
Ni$_2$MnAl, Ni$_2$MnGe and Co$_2$MnGa \cite{Zayak-PRB,Thomas}. In case of
Ni$_2$MnGa and Ni$_2$MnAl the acoustic phonon modes soften in
the range of $\zeta$
between $\approx$ 0.2-0.5 and 0.25-04, respectively.
This incommensurate softening is observed in experimental studies and
assigned to be at the origin of the modulations
\cite{Manosa-01PRB,Claudia}. In case of Ni$_2$MnGe the
softening is also present but the L2$_1$ structure of this crystal is not
stable which leads to additional instability at small values of $\zeta$.
In contrast to that, Co$_2$MnGe (Fig. \ref{comnge}) and Co$_2$MnGa do not show
an incommensurate phonon instability.

In order to compare how the softening appears in different Heusler alloys we
show in  Fig. \ref{modes} the calculated dispersions of the TA$_2$ phonon
modes of all five crystals. We emphazise here the fact that the position of
the softening can be related to the $e/a$ ratio of the corresponding Heusler
alloy.

Figure \ref{dos} shows Ga and Ge site projected DOS of Ni$_2$MnGa and
Co$_2$MnGe, respectively. We note here that the main origin
for the phonon instability can be related to nesting of the minority-spin
states close to the Fermi level \cite{Harmon}. The stable systems do not show
corresponding nesting. However, our calculations show additional features 
primarily related to the origin of the Fermi--surface nesting. Namely, in all
unstable Heusler alloys we find hybridization of unomalously low-lying optical
and acoustic phonons. We expect that this feature leads to an
additional strengthening of the electron-phonon coupling because of
localized mode--electron coupling.
In particular, this coupling might enhance the interaction of the
Ni-$d$ and Ga-$p$ minority-spin states. The electron-phonon
interaction will flatten corresponding
$d$ bands of Ni right at the Fermi level leading to a nesting shape of the
Fermi surface allowing for structural instrabilities and the formation of
martensitic modulated structures.   
In conclusion we can therefore say that the main driving force for martensitic
transformations in MSM alloys is still at debate and not a
completely solved question.

%%%%%%%%%%%%%%%%%%%%%%%%%%%%%%%%%%%%%%%%%%%%%%%%
\section{Acknowledgements}
Financial support by the German Science Council
(GRK 277 and SFB 491) is very much acknowledged.
%%%%%%%%%%%%%%%%%%%%%%%%%%%%%%%%%%%%%%%%%%%%%%%%

%\appendix 
%\section{}

%  \bibliographystyle{elsart-num}
%  \bibliography{zayak}

\end{document}